\documentclass[aps,prb,amsmath,twocolumn,superscriptaddress,showpacs]{revtex4}
\usepackage{amsmath}
\usepackage{amssymb}

\usepackage{graphicx}
\usepackage{bm} %Uncomment for using bold math
\usepackage{url}

\DeclareMathOperator{\sgn}{sgn}

\DeclareMathOperator{\Sp}{\mathrm{Sp}}

\begin{document}

\author{Mikhail S. Kalenkov}
\affiliation{I.E. Tamm Department of Theoretical Physics, P.N. Lebedev Physical 
Institute, 119991 Moscow, Russia}
\affiliation{Laboratory of Cryogenic Nanoelectronics, Nizhny Novgorod State 
Technical University, 603950 Nizhny Novgorod, Russia}
\author{Andrei D. Zaikin}
\affiliation{Institut f\"ur Nanotechnologie, Karlsruher Institut f\"ur 
Technologie (KIT), 76021 Karlsruhe, Germany}
\affiliation{I.E. Tamm Department of Theoretical Physics, P.N. Lebedev Physical 
Institute, 119991 Moscow, Russia}

%\author{Mikhail~S~Kalenkov$^{1,2}$, Andrei~D~Zaikin$^{3,1}$}

%\address{$^{1}$I.E. Tamm Department of Theoretical Physics, P.N. 
%Lebedev Physical Institute, 119991 Moscow, Russia}
%\address{$^{2}$Laboratory of Cryogenic Nanoelectronics, Nizhny Novgorod State 
%Technical University, 603950 Nizhny Novgorod, Russia}
%\address{$^{3}$Institut f\"ur Nanotechnologie, Karlsruher Institut f\"ur 
%Technologie (KIT), 76021 Karlsruhe, Germany}
%\ead{kalenkov@lpi.ru}

\title%[Enhancement of thermoelectric effect in diffusive superconducting bilayers]%
{Enhancement of thermoelectric effect in diffusive superconducting bilayers with magnetic interfaces}

\begin{abstract}
We demonstrate that thermoelectric currents in superconducting bilayers with a spin-active interface are controlled by the two competing processes. On one hand, spin-sensitive quasiparticle scattering at such interface generates electron-hole imbalance and yields orders-of-magnitude enhancement of the thermoelectric effect in the system. On the other hand, this electron-hole imbalance gets suppressed in the superconductor bulk due to electron scattering on non-magnetic impurities. As a result, large thermoelectric currents can only flow in the vicinity of the spin-active interface and
decay away from this interface at a distance exceeding the electron elastic mean free path $\ell$. The magnitude of the thermoelectric effect reaches its maximum provided $\ell$ becomes of order of the total bilayer thickness.
\end{abstract}

\pacs{74.25.fg, 74.45.+c, 74.78.Fk}

%\noindent{\it Keywords\/}:
%thermoelectric effect, electron-hole asymmetry, spin-active interface

\maketitle
%\ioptwocol

\section{Introduction}

Thermoelectric effect in superconductors attracts a lot of attention over last decades \cite{NL}.
Several earlier experiments \cite{Zavaritskii74,Falco76,Harlingen80} demonstrated that thermoelectric currents
flowing in superconductors in the presence of a non-zero temperature gradient can reach values exceeding the standard
theoretical predictions \cite{Galperin73}
by several orders of magnitude. While more experimental research is definitely needed to clarify the situation,
on a theory side there has been a substantial progress allowing to pinpoint the basic physical reason
that may yield a dramatic increase of the thermoelectric effect in superconducting compounds.
It was argued by a number of authors that such an increase can be observed provided electron-hole symmetry is
violated in the system. In this case thermoelectric currents do not anymore depend on a small
parameter $T/\varepsilon_F \ll 1$ (where $\varepsilon_F$ is the Fermi energy)
and may reach values as high
as the critical (depairing) current of a superconductor.

Theoretical models describing this physical situation are diverse embracing, e.g.,
conventional superconductors doped by magnetic impurities \cite{Kalenkov12},
unconventional superconductors with non-magnetic impurities \cite{LF},
superconductor-ferromagnet hybrid structures with the density of states spin-split by the
exchange and/or Zeeman fields \cite{Machon,Ozaeta} as well as various realizations of
superconducting-normal (SN) hybrids \cite{Seviour00,VP,KZ14} and superconducting bilayers \cite{mism}.
In particular, one can consider
an SN bilayer with a spin-active interface separating the two metals (see Fig. 1). Recently
we demonstrated \cite{KZ14} that interface scattering rates for electrons and holes in
such structures in general differ from each other thus providing a transparent physical
mechanism for the electron-hole imbalance generation. The latter, in turn, results in
huge thermoelectric currents flowing along the SN interface provided the left and the right
ends of the bilayer are maintained at different temperatures.

For the sake of simplicity in \cite{KZ14,mism} the limit of sufficiently clean metals was considered,
in which case electrons and holes move ballistically and scatter only at the SN interface.
In realistic metallic structures, however, quasiparticles may also scatter on non-magnetic
impurities in the bulk of the sample, on various boundary imperfections and so on. As a result,
quasiparticle motion inside a metal becomes diffusive rather than ballistic and the whole analysis \cite{KZ14,mism}
needs to be modified in order to account for a non-trivial interplay between
non-magnetic impurity scattering  and electron reflection at the spin-active SN interface.
Investigation of the electron-hole imbalance and the thermoelectric effect under such
conditions is the primary goal of our present work.

The structure of the paper is as follows. In section 2 we describe our quasiclassical formalism
which is then employed in section 3 in order to quantitatively analyze the effect of electron scattering on non-magnetic impurities in our system. In section 4 we evaluate the thermoelectric current in disordered superconducting bilayers with spin-active intermetallic interfaces and present a brief discussion of our results. Some technical details of our calculation are displayed in Appendix.

\begin{figure}
\centerline{ \includegraphics[width=80mm]{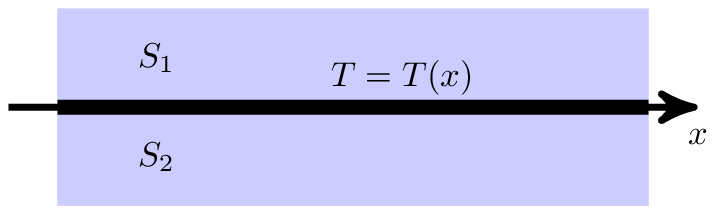} }
\caption{A metallic bilayer which consists of two superconductors $S_1$ and $S_2$ separated by a spin-active interface. The temperature $T(x)$ changes only in the direction parallel to the interface.}
\label{sfs-fig}
\end{figure}

\section{Quasiclassical formalism}
In what follows we will employ the quasiclassical theory of superconductivity. Within this theory the
electric current density $\bm{j}$ in the system can be expressed in terms of the Keldysh component of
the quasiclassical Green function
\begin{equation}
\bm{j}(\bm{r})= -\dfrac{e N_0}{8} \int d \varepsilon
\left< \bm{v}_F \Sp [\hat \tau_3 \hat g^K(\bm{p}_F, \bm{r},
\varepsilon) ] \right>,
\label{current}
\end{equation}
where $\bm{p}_F=m\bm{v}_F$ is the electron Fermi momentum vector, $\hat\tau_3$
is the Pauli matrix in the Nambu space, $N_0$ is the normal density of states at
the Fermi level, angular brackets $\left<\cdots\right>$ denote averaging over
the Fermi momentum directions and $\hat g^K$ is the  Keldysh block of the full
Green-Keldysh matrix
\begin{equation}
\check g =
\begin{pmatrix}
\hat g^R & \hat g^K \\
0 & \hat g^A
\end{pmatrix},
\label{green}
\end{equation}
Here and below the ``check'' symbol denotes the $8\times8$ matrices in
the Keldysh$\otimes$Nambu$\otimes$Spin space whereas the ``hat'' symbol implies the $4\times4$
matrices in the Nambu$\otimes$Spin space.
The matrix function $\check g$ obeys the transport-like Eilenberger equations \cite{bel}
\begin{equation}
\left[ \varepsilon \hat\tau_3 - \check\Delta(\bm{r}) - \check \sigma_{\text{imp}}, \check g \right]
+
i\bm{v}_F \nabla \check g (\bm{p}_F, \bm{r}, \varepsilon) =0
\label{Eil2}
\end{equation}
together with the normalization condition
\begin{equation}
\check g^2 =1.
\end{equation}
Here the self-energy $\check
\sigma_{\text{imp}}$ accounts for elastic electron
scattering off non-magnetic isotropic impurities randomly distributed in our sample.
It has the standard form
\begin{equation}
\check \sigma_{\text{imp}}
=
-i\dfrac{v_F}{2\ell} \left< \check g\right>,
\end{equation}
where $\ell$ is the electron elastic mean free path. The order parameter matrix $\check \Delta$
contains non-vanishing retarded and advanced components
\begin{equation}
\hat \Delta^A = \hat \Delta^R
=
\begin{pmatrix}
0 & \Delta \sigma_0 \\
-\Delta^* \sigma_0 & 0
\end{pmatrix},
\quad
\hat \Delta^K =0,
\end{equation}
where $\sigma_0$ is the unity matrix in the spin space and $\Delta$ is the superconducting order parameter
which will be chosen real throughout our consideration.

\begin{figure}
\centerline{ \includegraphics[width=80mm]{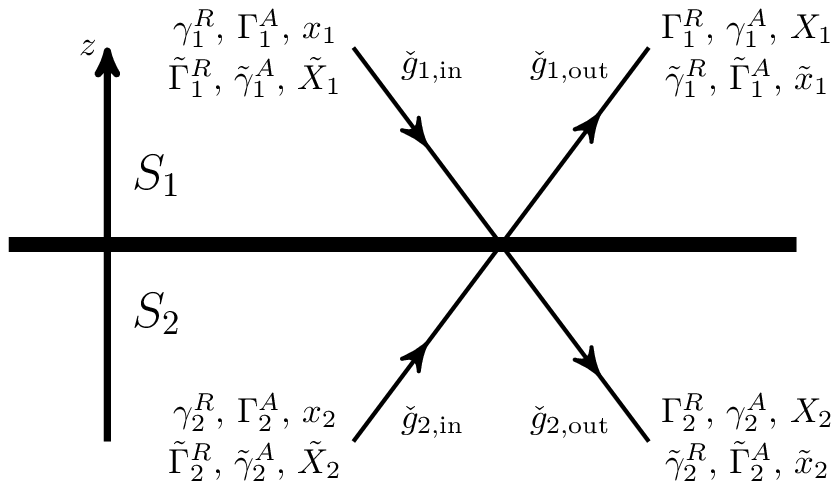} }
\caption{Boundary conditions matching the Green functions for incoming and outgoing momentum directions on both sides of the interface.}
\label{sis-fig2}
\end{figure}

The above quasiclassical equations should be supplemented by the proper boundary conditions
matching Green functions for incoming and outgoing momentum directions on both
sides of the interface (see Fig. \ref{sis-fig2}). Similarly to our earlier works \cite{KZ14,mism}, here we
will also assume that the two metals forming a bilayer are separated by a spin-active interface provided, e.g., by a thin ferromagnetic layer. The corresponding
boundary conditions for the quasiclassical propagators at such interfaces were formulated in Ref. \cite{Millis88}. Below we will use an equivalent approach
developed in Ref. \cite{Zhao04}.

We will stick to a simple model describing spin-dependent electron scattering at the interface. This model involves three parameters, i.e. the two interface transmission probabilities
$D_{\uparrow}$ and $D_{\downarrow}$ describing opposite spin directions as well as the so-called
spin mixing angle $\theta$ which is just the difference between the scattering phase shifts for the spin-up and spin-down electrons. Within this model the elements of the interface S-matrix take the form
\begin{gather}
S_{11}=S_{22}=
\sqrt{R_{\sigma}}e^{i\theta_{\sigma}/2},
\\
S_{12}=S_{21}=
i \sqrt{D_{\sigma}}e^{i\theta_{\sigma}/2},
\\
\underline{S}_{11}=\underline{S}_{22}=
\sqrt{R_{-\sigma}}e^{-i\theta_{\sigma}/2},
\\
\underline{S}_{12}=\underline{S}_{21}=
i \sqrt{D_{-\sigma}}e^{-i\theta_{\sigma}/2},
\end{gather}
where $\theta_{\sigma}=\theta \sigma_3$ is the $2\times 2$ diagonal matrix in the spin space.
The matrices $R_{\pm\sigma}$, $D_{\pm\sigma}$ are defined as
\begin{gather}
R_{\sigma}=
\begin{pmatrix}
R_{\uparrow} & 0 \\
0 & R_{\downarrow}
\end{pmatrix},
\quad
R_{-\sigma}=
\begin{pmatrix}
R_{\downarrow} & 0 \\
0 & R_{\uparrow}
\end{pmatrix},
\\
D_{\sigma}=
\begin{pmatrix}
D_{\uparrow} & 0 \\
0 & D_{\downarrow}
\end{pmatrix},
\quad
D_{-\sigma}=
\begin{pmatrix}
D_{\downarrow} & 0 \\
0 & D_{\uparrow}
\end{pmatrix},
\end{gather}
where $R_{\uparrow}=1-D_{\uparrow}$ and $R_{\downarrow}=1-D_{\downarrow}$ are the electron
reflection probabilities for the corresponding spin direction.
The above matrices constitute the building blocks of the full S-matrix for electrons
\begin{equation}
\mathcal{S}
=
\begin{pmatrix}
S_{11} & S_{12} \\
S_{21} & S_{22}
\end{pmatrix},
\end{equation}
and holes
\begin{equation}
\underline{\mathcal{S}}
=
\begin{pmatrix}
\underline{S}_{11} & \underline{S}_{12} \\
\underline{S}_{21} & \underline{S}_{22}
\end{pmatrix}.
\end{equation}
In what follows it will be convenient for us to make use
of the so-called Riccati parameterization \cite{Schopohl95,Eschrig00} for the Green functions involving four Riccati amplitudes and
two distribution functions (see appendix \ref{app-riccati} for more details).
Following \cite{Eschrig00} we denote the distribution functions $x$ and the Riccati amplitudes $\gamma$ by
the upper-case and the lower-case letters depending on the quasiparticle Fermi momentum (see Fig. \ref{sis-fig2}):
\begin{gather}
\check g_{i, \text{in}} = \check g_{i, \text{in}}
[\gamma_i^R, \tilde \Gamma_i^R, \Gamma_i^A, \tilde \gamma_i^A, x_i, \tilde X_i],
\quad i=1,2,
\\
\check g_{i, \text{out}} = \check g_{i, \text{out}}
[\Gamma_i^R, \tilde \gamma_i^R, \gamma_i^A, \tilde \Gamma_i^A, X_i, \tilde x_i],
\quad i=1,2,
\end{gather}
where the Riccati amplitudes $\gamma$, $\tilde \gamma_i$, $\Gamma_i$, $\tilde \Gamma_i$ and the distribution
functions $x_i$, $\tilde x_i$, $X_i$, $\tilde X_i$ are all $2\times 2$ matrices in the spin space.
The boundary conditions at the spin-active interface \cite{Zhao04} express the interface values of the ``upper-case''
functions $\Gamma$ and $X$ in terms of the ``lower-case'' ones $\gamma$ and $x$.

\section{Effect of impurity scattering}

Let us now investigate the effect of impurity scattering in superconducting bilayers consisting of two superconductors S$_1$ and S$_2$ situated respectively in half-spaces $z>0$ and $z<0$ and separated by a spin-active interface. All quantities will be labeled by the index 1 or 2 depending on whether they belong to the first or the second superconductor.

We will assume that the temperature $T=T(x)$ varies slowly as a function of the coordinate $x$ along the interface and does not depend on the coordinates $y$ and $z$.
Then the Keldysh Green function can be written in the form
\begin{equation}
\hat g^K_i =
\left[\hat g^R_i - \hat g^A_i\right]
\tanh\dfrac{\varepsilon}{2T(x)}
+
\hat g^a_i,
\end{equation}
where the term $\hat g^a_i$ is proportional to the temperature gradient $\partial_x T$ and, hence, remains sufficiently small.
The Green function $\hat g^a_i$ for incoming and outgoing momentum directions can be parameterized by the two distribution functions
\begin{gather}
\hat g_{i,in}^a =
2
\dfrac{
\begin{pmatrix}
x^a_i - \gamma^R_i \tilde X^a_i \tilde \gamma^A_i &
x^a_i \Gamma^A_i - \gamma^R_i \tilde X^a_i \\
\tilde X^a_i \tilde \gamma^A_i - \tilde \Gamma^R_i x^a_i &
\tilde X^a_i - \tilde \Gamma^R_i x^a_i \Gamma^A_i
\end{pmatrix}}
{
\left[1 - \gamma^R_i \tilde \Gamma^R_i \right]
\left[1 - \tilde \gamma^A_i \Gamma^A_i \right]
},
\\
\hat g_{i,out}^a =
2
\dfrac{
\begin{pmatrix}
X^a_i - \Gamma^R_i \tilde x^a_i \tilde \Gamma^A_i &
X^a_i \gamma^A_i - \Gamma^R_i \tilde x^a_i \\
\tilde x^a_i \tilde \Gamma^A_i - \tilde \gamma^R_i X^a_i &
\tilde x^a_i - \tilde \gamma^R_i X^a_i \gamma^A_i
\end{pmatrix}}
{
\left[1 - \gamma^R_i \tilde \Gamma^R_i \right]
\left[1 - \tilde \gamma^A_i \Gamma^A_i \right]
}.
\end{gather}
Here we already made use of the fact that within our model all matrices are diagonal in the spin space.
The functions $x^a_i$, $\tilde x^a_i$, $X^a_i$, $\tilde X^a_i$ are non-equilibrium parts of the distribution functions, i.e.
\begin{gather}
x_i= (1-\gamma^R_i \tilde \gamma_i^A) \tanh\dfrac{\varepsilon}{2T(x)} + x_i^a,
\\
\tilde x_i= - (1-\tilde \gamma^R_i \gamma_i^A) \tanh\dfrac{\varepsilon}{2T(x)} + \tilde  x_i^a,
\\
X_i= (1-\Gamma^R_i \tilde \Gamma_i^A) \tanh\dfrac{\varepsilon}{2T(x)} + X_i^a,
\\
\tilde X_i= - (1-\tilde \Gamma^R_i \Gamma_i^A) \tanh\dfrac{\varepsilon}{2T(x)} + \tilde X_i^a.
\end{gather}

As our final goal is to evaluate the thermoelectric current flowing along the interface between two superconductors, it is instructive to obtain the expression for the corresponding combination which enters into Eq. \eqref{current}, i.e.
\begin{equation}
\Sp (\hat \tau_3 \hat g^a_{i,\text{in}} + \hat \tau_3 \hat g^a_{i,\text{out}})
=
2\Sp
\left[
\dfrac{(X_i^a- \tilde X_i^a) (1+\gamma_i^R \gamma_i^A)}{(1 - \gamma_i^R \Gamma_i^R)(1 - \gamma_i^A \Gamma_i^A)}
\right] .
\label{gin-gout}
\end{equation}
Here we already employed the condition $x_i^a =
\tilde x_i^a$ satisfied within the linear response approximation we are going to use. The combination \eqref{gin-gout} contains a small factor $X_i^a- \tilde X_i^a$ proportional to the temperature gradient $\partial_x T$. This observation enables us to evaluate the Riccati amplitudes in Eq. \eqref{gin-gout} in thermodynamic equilibrium.

With the aid of the boundary conditions one can establish the relations between the interface values of the Riccati amplitudes. They read
%\begin{widetext}
\begin{multline}
\Gamma^R_1(0)=
\bigl[
\gamma^R_1(0)\sqrt{R_{\uparrow} R_{\downarrow}}
+
\gamma^R_2(0)\sqrt{D_{\uparrow} D_{\downarrow}}
-\\-
\gamma^R_1(0) (\gamma^R_2(0))^2 e^{i\theta_{\sigma}}
\bigr]
\bigl[
1-
(\gamma^R_2(0))^2 \sqrt{R_{\uparrow} R_{\downarrow}} e^{i\theta_{\sigma}}
-\\-
\gamma^R_2(0) \gamma^R_1(0) \sqrt{D_{\uparrow} D_{\downarrow}} e^{i\theta_{\sigma}}
\bigr]^{-1}
e^{i\theta_{\sigma}},
\end{multline}
\begin{multline}
\Gamma^A_1(0)=
\bigl[
\gamma^A_1(0)\sqrt{R_{\uparrow} R_{\downarrow}}
+
\gamma^A_2(0)\sqrt{D_{\uparrow} D_{\downarrow}}
-\\-
\gamma^A_1(0) (\gamma^A_2(0))^2 e^{-i\theta_{\sigma}}
\bigr]
\bigl[
1-
(\gamma^A_2(0))^2 \sqrt{R_{\uparrow} R_{\downarrow}} e^{-i\theta_{\sigma}}
-\\-
\gamma^A_2(0) \gamma^A_1(0) \sqrt{D_{\uparrow} D_{\downarrow}} e^{-i\theta_{\sigma}}
\bigr]^{-1}
e^{-i\theta_{\sigma}}.
\end{multline}
%\end{widetext}
The analogous expressions for $\Gamma^{R,A}_2(0)$ are derived from the above equations
by means of a trivial index replacement $1\leftrightarrow 2$. In the equilibrium
Riccati amplitudes depend on energy $\varepsilon$, momentum $\bm{p}_F$ and
coordinate $z$. Note that for brevity we do not indicate explicitly
the dependence of Riccati amplitudes on the energy $\varepsilon$ and momentum $\bm{p}_F$ arguments. We also note that the equations for the tilde Riccati amplitudes are redundant because of the
identities $\tilde \gamma^{R,A}_i = \gamma^{R,A}_i$ and $\tilde \Gamma^{R,A}_i =
\Gamma^{R,A}_i$ which remain applicable as long as the superconducting order parameter is chosen real.

With the aid of the above quasiclassical equations it is straightforward to demonstrate that the difference $X_i^a - \tilde X_i^a$ obeys the equation
\begin{multline}
i|v_{z}|(\sgn z)  \partial_z (X^a_i - \tilde X^a_i)
+\\+
\left(\tilde \varepsilon^R_i - \tilde \varepsilon^A_i -
\tilde \Delta^R_i \Gamma^R_i + \tilde \Delta^A_i \Gamma^A_i
\right) (X^a_i - \tilde X^a_i) =0,
\label{dxs}
\end{multline}
where $\tilde \varepsilon^{R,A}_i$ and $\tilde \Delta^{R,A}_i$ are respectively the renormalized energy and the order parameter defined as
\begin{equation}
\begin{pmatrix}
\tilde \varepsilon^{R,A}_i & \tilde \Delta^{R,A}_i \\
-\tilde \Delta^{R,A}_i & - \tilde \varepsilon^{R,A}_i
\end{pmatrix}
=
\begin{pmatrix}
\varepsilon & \Delta_i \\
-\Delta_i & -\varepsilon
\end{pmatrix}
-\hat \sigma_{i,\text{imp}}^{R,A}.
\end{equation}
Eq. \eqref{dxs} can easily be resolved with the result
\begin{multline}
X^a_i(z) - \tilde X^a_i(z)
=
\dfrac{
[1 - \gamma^R_i(z) \Gamma^R_i(z)][1 - \gamma^A_i(z) \Gamma^A_i(z)]
}{
[1 - \gamma^R_i(0) \Gamma^R_i(0)][1 - \gamma^A_i(0) \Gamma^A_i(0)]
}
\times\\\times
\left[ X^a_i(0) - \tilde X^a_i(0) \right]
\exp\left(-
\dfrac{2\sgn z}{|v_{z}|}
\int^z_{0} w_i(z')dz'
\right),
\end{multline}
where
\begin{equation}
2i w_i = \tilde \varepsilon^R_i - \tilde \varepsilon^A_i -
\tilde \Delta^R_i \gamma^R_i + \tilde \Delta^A_i \gamma^A_i,
\end{equation}
Exploiting the boundary conditions we can express the difference $X^a_i(0) - \tilde X^a_i(0)$ at the interface in terms of the interface values $x^a_i(0)$,
\begin{multline}
X_1^a(0) - \tilde X_1^a(0)
=(R_{\uparrow} - R_{\downarrow})
\left[1-\gamma^R_2(0)\gamma^A_2(0)\right]
\sigma_3
\times\\\times
\dfrac{\left[1+\gamma^R_2(0)\gamma^A_2(0)\right] x_1^a(0)
-
\left[1+\gamma^R_1(0)\gamma^A_1(0)\right] x_2^a(0)
}{
\left|1-
(\gamma^R_2(0))^2 \sqrt{R_{\uparrow} R_{\downarrow}} e^{i\theta_{\sigma}}
-
\gamma^R_2(0) \gamma^R_1(0) \sqrt{D_{\uparrow} D_{\downarrow}} e^{i\theta_{\sigma}}\right|^2
},
\end{multline}
and similarly for $X_2^a(0) - \tilde X_2^a(0)$.

The interface value $x_i^a(0)$ is recovered from the equation
\begin{gather}
2|v_{z}| \sgn z \partial_z x^a_i
+
w_i x^a_i =
-
v_{x}\dfrac{\varepsilon (1- \gamma^R_i \gamma^A_i )}{ T^2 \cosh^2(\varepsilon/2T)}
\partial_x T  ,
\end{gather}
which yields
\begin{multline}
x^a_1(0)=
\dfrac{v_{x}\varepsilon \partial_x T}{2|v_{z}|T^2 \cosh^2(\varepsilon/2T)}
\times\\\times
\int\limits^{\infty}_0
\left[1-\gamma_1^R(z)\gamma_1^A(z)\right]
\exp\left(-
\dfrac{2}{|v_{z}|}
\int^z_{0} w_1(z')dz'
\right)dz.
\end{multline}
An analogous expression can be established for $x^a_2(0)$.

Introducing the characteristic lengths $L_i^{\pm}$ defined by means of the equations
\begin{multline}
\int\limits_0^{\infty}
[1\pm\gamma^R_1(z) \gamma^A_1(z)]
\exp\left(-
\dfrac{2}{|v_{z}|}
\int^z_{0} w_1(z')dz'
\right) dz
=\\=
\dfrac{|v_{z}|}{v_F}
[1\pm\gamma^R_1(0) \gamma^A_1(0)]L_1^{\pm},
\end{multline}
and
\begin{multline}
\int\limits_{-\infty}^0
[1\pm\gamma^R_2(z) \gamma^A_2(z)]
\exp\left(-
\dfrac{2}{|v_{z}|}
\int^0_z w_2(z')dz'
\right) dz
=\\=
\dfrac{|v_{z}|}{v_F}
[1\pm\gamma^R_2(0) \gamma^A_2(0)]L_2^{\pm},
\end{multline}
one can conveniently rewrite the interface values $x^a_i(0)$ in a compact form
\begin{gather}
x^a_i(0)=
\dfrac{v_{x}\left[1-\gamma_i^R(0)\gamma_i^A(0)\right]
\varepsilon L_i^- }{2 v_F T^2 \cosh^2(\varepsilon/2T)}  \partial_x T.
\end{gather}
The above equations allow to fully describe the effect of electron scattering on non-magnetic impurities and to evaluate the thermoelectric currents in the system under consideration.

\section{Thermoelectric currents}
Combining the results derived in the previous section, from Eq. \eqref{current} we obtain the expression for the current density, e.g., in the superconductor $S_1$ ($z>0$). It reads
%\begin{widetext}
\begin{multline}
j_1(z)=
-\dfrac{e N_0}{8v_F} \partial_x T
\int
\dfrac{\varepsilon  d \varepsilon}{T^2 \cosh^2(\varepsilon/2T)}
\Biggl<
v_{x}^2
\Theta (-v_{z}) 
\times\\\times
(R_{\uparrow} - R_{\downarrow})
\Sp \Biggl\{
\sigma_3\mathcal{A}_1^+(z)
\exp\left(-
\dfrac{2}{|v_{z}|}
\int^z_{0} w_1(z')dz'
\right)
\times\\\times
\mathcal{A}_2^-(0)
\left[
\mathcal{A}_2^+(0)
\mathcal{A}_1^-(0)
L_1^-
-
\mathcal{A}_1^+(0)
\mathcal{A}_2^-(0)
L_2^-\right]
\mathcal{N}
\Biggr\}
\Biggr>,
\label{thermoIS}
\end{multline}
%\end{widetext}
where 
\begin{multline}
\mathcal{N}=\bigl|1-
(\gamma^R_1(0))^2 \sqrt{R_{\uparrow} R_{\downarrow}} e^{i\theta_{\sigma}}
-
(\gamma^R_2(0))^2 \sqrt{R_{\uparrow} R_{\downarrow}} e^{i\theta_{\sigma}}
-\\-
2\gamma^R_2(0) \gamma^R_1(0) \sqrt{D_{\uparrow} D_{\downarrow}} e^{i\theta_{\sigma}}
+
(\gamma^R_2(0) \gamma^R_1(0))^2 e^{2i\theta_{\sigma}}
\bigr|^{-2},
\end{multline}
\begin{equation}
\mathcal{A}_i^{\pm}(z)=1\pm\gamma^R_i(z)\gamma^A_i(z),
\end{equation}
and $\Theta (y)$ is the Heavyside step function. The current density in the superconductor $S_2$ ($z<0$) is
obtained from Eq. \eqref{thermoIS} by replacing $1\leftrightarrow 2$
and $\int_0^z \leftrightarrow \int^0_z$. From these results we observe that thermoelectric currents on two sides of the interface have opposite signs, i.e.
these currents can flow in {\it opposite directions}.

It also follows from the above results that impurity scattering leads
to the exponential decay of the current density far from the spin-active
interface. The characteristic length of the decay is controlled by function
$w_i$ and depends both on the electron energy and on its momentum direction.
Far from the interface the function $w_i$ can easily be established analytically
since in this limit the retarded and advanced Green functions tend to their bulk values. After a simple calculation one finds
\begin{equation}
w_i (\varepsilon)=
\begin{cases}
\dfrac{v_F}{2\ell_i}, &|\varepsilon| > \Delta_i,
\\
\dfrac{v_F}{2\ell_i} + \sqrt{\Delta_i^2 - \varepsilon^2} , &|\varepsilon| <
\Delta_i,
\end{cases}
\end{equation}
This result implies that the thermoelectric current is confined to the interface and decays deep into the superconducting bulk at a typical length not exceeding the corresponding elastic mean free path $\ell_{1(2)}$.

Integrating Eq. \eqref{thermoIS} over $z$
we obtain the net thermoelectric current flowing along the interface
%\begin{widetext}
\begin{multline}
I=
-\dfrac{e N_0}{8v_F^2} \partial_x T
\int
\dfrac{\varepsilon  d \varepsilon}{T^2 \cosh^2(\varepsilon/2T)}
\biggl<
v_{f,x}^2 |v_{z}|
\Theta (-v_{z}) 
\times\\\times
(R_{\uparrow} - R_{\downarrow})
\Sp \biggl\{
\sigma_3
\bigl[
\mathcal{A}_2^-(0)
\mathcal{A}_1^+(0)
L_1^+
-
\mathcal{A}_1^-(0)
\mathcal{A}_2^+(0)
L_2^+
\bigr]
\times\\\times
\left[
\mathcal{A}_2^+(0)
\mathcal{A}_1^-(0)
L_1^-
-
\mathcal{A}_1^+(0)
\mathcal{A}_2^-(0)
L_2^-\right]
\mathcal{N}
\biggr\}
\biggr>.
\label{thermoIdiff}
\end{multline}
%\end{widetext}
Just as in the ballistic limit, the above expression for the thermoelectric current becomes zero if the interface
transmission probabilities for the opposite spin directions coincide $D_{\uparrow}= D_{\downarrow}$ and/or the spin mixing angle $\theta$ equals to zero. In addition, the current density (\ref{thermoIS}) and, hence, also the total current (\ref{thermoIdiff}) vanish for identical superconductors S$_1$ and S$_2$
(in this case one has $\gamma^{R(A)}_1(0)=\gamma^{R(A)}_2(0)$ and $L_1^{\pm}=L_2^{\pm}$) indicating the absence of the electron-hole asymmetry in this specific limit.

Provided $D_{\uparrow}\neq D_{\downarrow}$, $\theta \neq 0$ and the superconductors
S$_1$ and S$_2$ are not identical (one of them can also be a normal metal) the current (\ref{thermoIdiff}) does not vanish and under certain conditions can reach values orders of magnitude higher than, e.g., in normal metals. The exact evaluation of
Eq. (\ref{thermoIdiff}) in a general case can only be performed numerically. However,
simple estimates can be obtained in certain limits.

For instance, in the tunneling limit $D_{\uparrow}, D_{\downarrow} \ll 1$ and for the case of diffusive superconductors with very different mean free path values (i.e. for $\ell_1^2 + \ell_2^2 \gg \ell_1
  \ell_2$) the expression (\ref{thermoIdiff}) reduces to a much simpler form:
\begin{multline}
I=
\dfrac{e N_0}{8v_F^2} \partial_x T
\int
\dfrac{(\ell_1^2+\ell_2^2) \varepsilon  d \varepsilon}{T^2
\cosh^2(\varepsilon/2T)}
\Bigl<
v_{x}^2 |v_{z}|
\Theta (-v_{z})
\times\\\times
(D_{\uparrow} - D_{\downarrow})
\left[\nu_{1\uparrow}(0) \nu_{2\uparrow}(0)-
\nu_{1\downarrow}(0) \nu_{2\downarrow}(0) \right]
\Bigr>,
\label{thermoIdiff2}
\end{multline}
where $\nu_{i\uparrow(\downarrow)}(0)$ are the momentum and energy resolved densities
of states at the interface for the opposite electron spin orientations.

At intermediate temperatures $T \sim \Delta$ we can roughly estimate the magnitude of the thermoelectric current as
\begin{equation}
I \sim e N_0 v_F\ell^2 (R_{\uparrow} - R_{\downarrow}) \sin\theta\partial_x T .
\label{estim}
\end{equation}
It is instructive to compare this result with the thermoelectric current $I_{\text{norm}}$ flowing in our bilayer in its normal state. Making use of the well known Mott relation for the thermoelectric coefficient of normal metals, from (\ref{estim}) we obtain
\begin{equation}
\dfrac{I}{I_{\text{norm}}}
\sim
\dfrac{\ell}{d}\dfrac{\varepsilon_F}{T_c}(R_{\uparrow} - R_{\downarrow})\sin\theta,
\label{est2}
\end{equation}
where $d$ is the total thickness of our bilayer, $T_c$ is the critical temperature of the bulk superconductor. Setting $\ell \sim d$, $R_{\uparrow} - R_{\downarrow} \sim 1$ and $\sin\theta \sim 1$ we immediately arrive at the conclusion that the thermoelectric current $I$ in the superconducting state can be enhanced by a very large factor up to $\varepsilon_F/T_c \gg 1$ as compared to that in the normal state $I_{\text{norm}}$.

Summarizing our results, we arrive at the following physical picture. Different scattering rates for electrons and holes at the spin-active interface result in
electron-hole imbalance generation \cite{KZ14} which in turn may yield to orders-of-magnitude enhancement of thermoelectric currents in our system. On the other hand, scattering on non-magnetic impurities tends to suppress this imbalance deep in the metal bulk. Hence, large thermoelectric currents can only flow in the vicinity of the interface. The characteristic current decay
length $L_i^{+}$ away from the interface turns out to be of order of the corresponding elastic mean free path. E.g., in the diffusive limit (i.e. provided $\ell_i$ remains shorter than the superconducting coherence length) one simply has $L_i^+ = L_i^- = \ell_i$. Accordingly, the magnitude of the thermoelectric current $I$ increases with
increasing $\ell_i$ and reaches its maximum when the elastic mean free becomes of order of the total bilayer thickness $d$, cf. Eq. (\ref{est2}). A similar trend was also observed within a different model of a superconductor doped by magnetic impurities \cite{Kalenkov12}.

Let us also note that although the general structure of Eq. \eqref{thermoIS} is quite similar to that of our earlier results \cite{KZ14,mism} derived for ballistic bilayers, it is not possible to directly recover the latter by setting $\ell_i\rightarrow \infty$ in the expression \eqref{thermoIS}. This is because our present results were derived
under the assumption about the existence of a local temperature $T(x)$ in our system slowly varying along the $x$-axis. Accordingly, Eq. \eqref{thermoIS} holds under the condition $\ell_i \lesssim T/|\partial_x T|$. No such condition was employed in the analysis \cite{KZ14,mism}.

Nevertheless, a formal replacement $L_i^- \partial_x T \rightarrow \Delta T$ (where
$\Delta T$ is the total temperature difference applied to our system) together with
putting $\ell_i$ equal to infinity makes the structure of Eq. \eqref{thermoIS} fully equivalent to that of the ballistic result \cite{KZ14,mism}. The latter observation implies that our main conclusion about the parametric enhancement of the thermoelectric effect in superconducting structures with spin-active interfaces is robust and is not
sensitive to the details of the adopted model.

MSK acknowledges support from Grant NNSTU No. 11.G34.31.0029 under Russian 
Governement Decree \#220.

\appendix
\section{Green function parameterization}
\label{app-riccati}
In the course of our analysis we employ the so-called Riccati parameterization \cite{Schopohl95,Eschrig00} for the retarded and advanced Green functions, i.e. we set
\begin{equation}
\hat g^{R,A}=\pm
    \hat N^{R,A}
    \begin{pmatrix}
    1+\gamma^{R,A} \tilde \gamma^{R,A} & 2\gamma^{R,A} \\
    -2 \tilde \gamma^{R,A} & -1- \tilde \gamma^{R,A}  \gamma^{R,A} \\
    \end{pmatrix},
    \label{graparam}
\end{equation}
where
\begin{equation}
\hat N^{R,A}=
    \begin{pmatrix}
    (1-\gamma^{R,A} \tilde \gamma^{R,A})^{-1} & 0 \\
    0 & (1-\tilde \gamma^{R,A}  \gamma^{R,A} )^{-1} \\
    \end{pmatrix}.
    \label{nrparam}
\end{equation}
and the Riccati amplitudes $\gamma^{R,A}$, $\tilde \gamma^{R,A}$ are $2\times2$ matrices in the spin space. The expression for the Keldysh Green function contains two distribution functions $x^K$, $\tilde x^K$ also being $2\times2$ matrices in spin space
\begin{equation}
\hat g^K=
2
\hat N^R
\begin{pmatrix}
x^K - \gamma^R  \tilde x^K  \tilde \gamma^A &
-\gamma^R  \tilde x^K + x^K  \gamma^A \\
-\tilde \gamma^R  x^K + \tilde x^K  \tilde \gamma^A &
\tilde x^K - \tilde \gamma^R  x^K  \gamma^A \\
\end{pmatrix}
\hat N^A.
\label{gkparam}
\end{equation}

The amplitudes $\gamma^{R,A}$, $\tilde \gamma^{R,A}$ obey the Riccati equation
\begin{gather}
i\bm{v}_F \nabla \gamma^{R,A}=
\begin{pmatrix}
1 & \gamma^{R,A}
\end{pmatrix}
\hat h^{R,A}
\begin{pmatrix}
-\gamma^{R,A} \\ 1
\end{pmatrix},
\\
i\bm{v}_F \nabla \tilde \gamma^{R,A}=
\begin{pmatrix}
\tilde \gamma^{R,A}& 1
\end{pmatrix}
\hat h^{R,A}
\begin{pmatrix}
1 \\  -\tilde \gamma^{R,A}
\end{pmatrix},
\end{gather}
and the distribution functions $x^K$ and $\tilde x^K$ satisfy the transport-like equations
\begin{multline}
i\bm{v}_F \nabla x^K
=
x^K
\begin{pmatrix}
1 & 0 \\
\end{pmatrix}
\hat h^A
\begin{pmatrix}
1 \\ -\tilde\gamma^A \\
\end{pmatrix}
+\\+
\begin{pmatrix}
1 & \gamma^R \\
\end{pmatrix}
\hat h^K
\begin{pmatrix}
1 \\ -\tilde \gamma^A \\
\end{pmatrix}
-
\begin{pmatrix}
1 & \gamma^R \\
\end{pmatrix}
\hat h^R
\begin{pmatrix}
1 \\ 0 \\
\end{pmatrix}
x^K,
\end{multline}
and
\begin{multline}
i\bm{v}_F \nabla \tilde x^K
=
\tilde x^K
\begin{pmatrix}
0 & 1 \\
\end{pmatrix}
\hat h^A
\begin{pmatrix}
-\gamma^A \\ 1 \\
\end{pmatrix}
-\\-
\begin{pmatrix}
\tilde \gamma^R & 1 \\
\end{pmatrix}
\hat h^K
\begin{pmatrix}
- \gamma^A \\ 1 \\
\end{pmatrix}
-
\begin{pmatrix}
\tilde \gamma^R & 1 \\
\end{pmatrix}
\hat h^R
\begin{pmatrix}
0 \\ 1 \\
\end{pmatrix}
\tilde x^K.
\end{multline}
Here the matrices $\hat h^{R,A,K}$ denote respectively the retarded, advanced and Keldysh components of the matrix $\check h= \varepsilon \hat\tau_3 - \check\Delta(\bm{r}) - \check \sigma_{\text{imp}}$.

\end{document}